\documentclass[a4paper, amsfonts, amssymb, amsmath, reprint, showkeys, nofootinbib, twoside,aps,prl]{revtex4-2}

\RequirePackage[T1]{fontenc}

\usepackage[english]{babel}
\usepackage[utf8]{inputenc}
\usepackage[colorinlistoftodos, color=green!40, prependcaption]{todonotes}
\usepackage{xcolor}
\usepackage{xspace}
\usepackage{lineno}
\usepackage{amssymb}
\usepackage{url}
\usepackage{ulem}
\usepackage{amsthm}
\usepackage{mathtools}
\usepackage{physics}
\usepackage{xcolor}
\usepackage{graphicx}
\usepackage[left=23mm,right=13mm,top=35mm,columnsep=15pt]{geometry} 
\usepackage{adjustbox}
\usepackage{placeins}
\usepackage[T1]{fontenc}
\usepackage{lipsum}
\usepackage{csquotes}
\usepackage[Symbolsmallscale]{upgreek}
\newcommand{\pp}           {pp\xspace}

\newcommand{\mt}           {\ensuremath{m_{\rm T}}\xspace}

\newcommand{\pP}           {\ensuremath{\mbox{pp}}\xspace}

\newcommand{\pL}           {\ensuremath{\rm \mbox{p}\Lambda}\xspace}

\newcommand{\Sig} {\ensuremath{\rm\Sigma}\xspace}

\newcommand{\nineH}        {$\sqrt{s}~=~0.9$~Te\kern-.1emV\xspace}
\newcommand{\seven}        {$\sqrt{s}~=~7$~Te\kern-.1emV\xspace}
\newcommand{\onethree}        {$\sqrt{s}~=~13$~Te\kern-.1emV\xspace}
\newcommand{\twoH}         {$\sqrt{s}~=~0.2$~Te\kern-.1emV\xspace}
\newcommand{\twosevensix}  {$\sqrt{s}~=~2.76$~Te\kern-.1emV\xspace}
\newcommand{\five}         {$\sqrt{s}~=~5.02$~Te\kern-.1emV\xspace}
\newcommand{\twosevensixnn}{$\sqrt{s_{\mathrm{NN}}}~=~2.76$~Te\kern-.1emV\xspace}
\newcommand{\fivenn}       {$\sqrt{s_{\mathrm{NN}}}~=~5.02$~Te\kern-.1emV\xspace}

\newcommand{\MeV}  {\ensuremath{\text{Me\kern-.1emV}}\xspace}
\newcommand{\MeVc}  {\ensuremath{\text{Me\kern-.1emV/}c}\xspace}
\newcommand{\MeVcc}  {\ensuremath{\text{Me\kern-.2emV/}c^2}\xspace}
\newcommand{\GeV}  {\ensuremath{\text{Ge\kern-.1emV}}\xspace}
\newcommand{\GeVc}  {\ensuremath{\text{Ge\kern-.1emV/}c}\xspace}
\newcommand{\GeVcc}  {\ensuremath{\text{Ge\kern-.2emV/}c^2}\xspace}
\newcommand{\TeV}  {\ensuremath{\text{Te\kern-.1emV}}\xspace}

\newcommand{\lmb}          {\ensuremath{\Lambda}\xspace}

\newcommand{\Casc}            {\ensuremath{\Xi}\xspace}

\newcommand{\SN}{\ensuremath{\rm N \Sigma}\xspace}
\newcommand{\LN}{\ensuremath{\rm N \Lambda}\xspace}
\newcommand{\LNN}{\ensuremath{\rm NN \Lambda}\xspace}

\newcommand{\Chieft}           {\ensuremath{\rm \chi EFT}\xspace}

\usepackage[pdftex, pdftitle={Article}, pdfauthor={Author}]{hyperref} 
\usepackage{microtype} 

\begin{document}
\title{Constraining the p$\Lambda$ interaction from a combined analysis of scattering data and correlation functions}

\author{D. L. Mihaylov$^{a,b}$}
\email{dimitar.mihaylov@mytum.de}
\author{J. Haidenbauer$^{c}$}
\email{j.haidenbauer@fz-juelich.de}
\author{V. Mantovani Sarti$^{a}$}
\email{valentina.mantovani-sarti@tum.de}

\affiliation{$^a$Technische Universit\"at M\"unchen, Physics Department, James-Franck-Str., 85748 Garching, Germany}
\affiliation{$^b$Sofia University, Faculty of Physics, 5 J. Bourchier Blvd, 1164 Sofia, Bulgaria}
\affiliation{$^c$Forschungszentrum J\"ulich, Institute for 
Advanced Simulation (IAS-4), 52428 J\"ulich, Germany}

\begin{abstract}

This work provides the first combined analysis of low-energy \pL scattering, considering both cross section and correlation data. The obtained results establish the most stringent constraints to date on the two-body \pL interaction, pointing to a weaker attraction than so far
accepted. 
The best set of scattering lengths for the spin singlet and triplet are found to range from $f_0, f_1 = (2.1, 1.56)$ to $(3.34, 1.18)~$fm. 
With a chiral NY potential fine-tuned to those scattering parameters, the
in-medium properties of the $\Lambda$ are explored and a potential
depth of $U_\Lambda= -36.3\pm 1.3 \mathrm{(stat)}^{+2.5}_{-6.2}\mathrm{(syst)}$~MeV
is found at nuclear matter saturation density. 
\end{abstract}

\maketitle

\noindent
\textit{Introduction:} The strong interaction between nucleons (N) and hyperons (Y=\lmb,\Sig,\Casc) plays a significant role in various aspects of nuclear and hadronic physics, ranging from the structure and properties of hypernuclei to the Equation of State (EoS)
of neutron stars (NS)~\cite{GalReview,TolosFabbiettiReview}. The recent observations of gravitational waves emitted from NS mergers~\cite{LIGOScientific2017,LIGOScientificMulti} and precise constraints on NS radii provided by the NICER collaboration~\cite{NICER1,NICER2,NICER3,NICER4} triggered a renewed interest in the presence of strange degrees of freedom in these compact objects~\cite{VidanaHyp,SchaffnerBielichBook}.\\
\indent
Specifically, the \LN system represents an important pillar for our understanding of the low-energy QCD dynamics between ordinary matter and strange particles~\cite{Povh:1981nz}. Chiral effective field theory (\Chieft)~\cite{Weinberg1,Haidenbauer:NLO13,Haidenbauer:NLO19,Haidenbauer:NNLO} is an excellent tool to study the \LN interaction, since it offers the
possibility to systematically improve the results by considering higher-order terms in the Lagrangian. These effective approaches rely on the availability of experimental data to determine the a priori unknown low-energy constants (LECs) associated with the contact interactions included in the Lagrangian. Until recently, the experimental constraints on the \LN interaction, and, in general, on the $S=-1$ baryon-baryon interaction, consisted primarily of scattering data~\cite{Sechi-Zorn:pLambda,Alexander:pLambda,Eisele:1971mk} and measurements of \lmb-hypernuclei binding energies~\cite{GalReview,Hashimoto:2006aw}. Elastic and inelastic cross section data, probing the transition $\LN \leftrightarrow \SN$, are relatively scarce and not available down to the threshold. Results from the COSY experiment delivered input on the $\LN-\SN$ dynamics by studying the $\mathrm{pp}\rightarrow \mathrm{pK}^+ \Lambda$~\cite{COSY-TOF:2013uqx,COSY-TOF:2016qxd}. New scattering measurements on both \LN and \SN were reported by CLAS~\cite{CLAS:2021gur} and E40~\cite{J-PARCE40:2021bgw,J-PARCE40:2021qxa,J-PARCE40:2022nvq} collaborations, adding constraints at higher momenta. Additionally, measured binding energies and lifetimes of light \lmb-hypernuclei \cite{GalReview} provide complementary information on the strength of the \LN interaction, for example with regard to the hypertriton~\cite{ALICE:HyperTriton,STAR:HyperTriton}  specifically on the singlet state \cite{Haidenbauer:NLO19}. \\
\indent
Data on hypernuclei in the medium and heavy mass regime have been employed to deduce the depth of the \lmb single-particle potential $U_\lmb$ in infinite nuclear matter at nuclear
saturation density $\rho_0=0.166$~fm$^{-1}$. An overall attraction of $\approx -30$ MeV is typically reported~\cite{Millener:UL1988,GalReview}, and this benchmark eventually serves as input for studies aiming to infer the behavior of \lmb hyperons at NS core densities, reaching few times $\rho_0$. 
Specifically, in investigations that commence with NY potentials describing N$\Lambda$ and N$\Sigma$ scattering data, the generally somewhat too attractive contribution from the two-body interaction is counterbalanced by an appropriate repulsive three-body \LNN component to meet the $\approx -30$ MeV constraint~\cite{Gerstung:2020ktv,Logoteta:2019utx}.\\
\indent
In the last years, novel data based on two-particle correlations,
involving strange hadrons, have become available and offer high-precision experimental insight into the $S=-1,-2,-3$ baryon-baryon interaction~\cite{Femtoreview,ALICE:pLCoupled, ALICE:2019buq, ALICE:2019hdt, ALICE:2020mfd, ALICE:2022uso,HADES:pL}. The recent measurement of the \pL correlation function in \pp collisions at \onethree by the ALICE Collaboration~\cite{ALICE:pLCoupled} provided the most precise data on this system down to threshold, accompanied by the first experimental observation of the opening of the coupled \SN channel in a two-body final state.
Ongoing and future experimental efforts are posed to advance our understanding of the \LN interaction and the $\LN \leftrightarrow \SN$ dynamics. These efforts include the utilization of polarized \lmb beams and high-precision hypernuclear spectroscopy~\cite{JPARCHEFEX_Third}, as well as  statistically improved correlation functions during the ongoing LHC Run 3 data taking~\cite{ALICERun3Run4}.\\
\indent
In this work, we, for the first time, perform a combined analysis of available \LN scattering and correlation data to constrain the corresponding \pL scattering parameters. 
The present study utilizes an Usmani-type potential and an interaction based on $\chi$EFT. 
Anticipating our main finding, the analysis suggests an overall less attractive \LN interaction compared to what has formed the basis for theoretical investigations so far. 
To explore the possible implications for the NS scenarios
\cite{Gerstung:2020ktv,Logoteta:2019utx},
appropriately re-adjusted chiral potentials are employed to evaluate the single-particle potential $U_\lmb$ at $\rho_0$. 

\newline\newline
\noindent
\textit{p$\Lambda$ interaction}: For the analysis of the \pL correlation data at low momenta, we employ wave functions
generated from two different types of potentials. 
To thoroughly explore the sensitivity of the effective range parameters to the measured \pL correlation functions, 
we use the Usmani potential~\cite{Bodmer:1984gc}, which includes spin dependence but lacks coupling to the N$\Sigma$ channel.
This allows for a realistic description of the N$\Lambda$ interaction below the N$\Sigma$ threshold, which is relevant for 
determining the effective range parameters.
The Usmani potential includes a Woods-Saxon-type repulsive core $V_C$, as well as a two-pion exchange tail constructed from a modified one-pion exchange tensor potential~\cite{Wang:1999bf}. 
The repulsive core reads: 
\begin{equation}\label{eq:UsmCore}
    V_C(r) = W_C\left[1+\mathrm{exp}\left(\frac{r-R_C}{d_C}\right)\right]^{-1}.
\end{equation}
The parameters $(W_C, R_C, d_C)$ are phenomenological in nature, and thus, we will adjust them in accordance with the analyzed data. The remaining parameters entering the Usmani potential 
are taken from~\cite{Wang:1999bf}.

In addition, we employ a modern NY potential derived within SU(3) $\chi$EFT \cite{Haidenbauer:NLO13,Haidenbauer:NLO19}, 
specifically, we utilize the NY potential NLO19
established in Ref.~\cite{Haidenbauer:NLO19}. 
This potential incorporates contributions up to 
next-to-leading order (NLO)  
in the chiral expansion, in particular it includes contributions from one- and 
two-pseudoscalar-meson exchange diagrams, involving the
Goldstone bosons $\pi$, $K$, $\eta$, 
and from four-baryon contact terms 
(without and with two derivatives), where the latter
encode the unresolved short-distance dynamics.
The LECs associated with these contact
terms are free parameters and  
have been established by a global fit to a set of
$36$ \pL and N$\Sigma$ low energy scattering data points \cite{Haidenbauer:NLO19}, available since the 1960s. 
SU(3) flavor symmetry has been imposed which reduces the number of independent contact terms or LECs, respectively.  
Then, in the two $S$-wave states $^1S_0$ and $^3S_0$, which dominate the scattering observables at low energies, 
there are 10 LECs \cite{Haidenbauer:NLO19}, with two of them inferred from the NN sector via the imposed SU(3) symmetry.

The present work incorporates the \pL correlation data measured by ALICE  
as an additional experimental constraint~\cite{ALICE:Source}.
To accommodate this, some of the LECs have to be varied 
in the search for the optimal strength of the N$\Lambda $ interaction.
Thereby, we aim at preserving the good description 
of the p$\Sigma^+$ and p$\Sigma^-$ data, as well as the 
p$\Sigma^-\to \mathrm{n}\Lambda$ transition cross 
section provided by the 
original potential~\cite{Haidenbauer:NLO19}.
The simplest and most efficient way to guarantee this is to relax the strict SU(3) symmetry for the contact interactions in the 
N$\Lambda$ and N$\Sigma$ forces \cite{Haidenbauer:NLO13,Haidenbauer:NLO19}. Therefore, 
following the procedure in \cite{Le:2019gjp}, 
we introduce an SU(3) symmetry breaking in the leading-order contact terms 
that contribute to the N$\Lambda $ interaction in the relevant $^1S_0$ and 
$^3S_1$ partial waves. 
We emphasize that such an SU(3) symmetry breaking at NLO 
is well in line with 
$\chi$EFT and the associated power counting 
\cite{Haidenbauer:NLO13,Petschauer:2013uua}. 
\newline\newline
\textit{Analysis:} The combined analysis of scattering and femtoscopic data is based on 12 data points for the \pL elastic cross section, as well as six \pL correlation functions measured in different ranges of pair transverse mass \mt in \pP collisions at 13~TeV. In particular, six points of the cross section stem from the work of Alexander et al.~\cite{Alexander:pLambda}, where we opt for those from the second choice of binning described in the corresponding work, while the remaining six data points are taken from the work of Sechi-Zorn et al.~\cite{Sechi-Zorn:pLambda}. The femtoscopy data originates from the ALICE measurement of the \pL correlation function in high-multiplicity (HM) \pp collisions at 13~TeV~\cite{ALICE:Source}. 
When analyzed below the N$\Sigma$ threshold, the $S$-waves are sufficient to account for the interaction. The parameters of the repulsive core $V_C(r)$ in Eq.~(\ref{eq:UsmCore}) are fitted independently for the spin singlet (S=0) and triplet (S=1) states, resulting in 
a total of 6 free parameters. 
The Usmani potential has been integrated into the CATS framework~\cite{Mihaylov:2018rva}, which is capable of evaluating the corresponding cross section and correlation function by solving the Schr\"odinger equation. While this is sufficient to describe the cross section data, femtoscopic data necessitates additional knowledge of the two-particle emission source $S(\mt,r)$, as demanded by the Koonin-Pratt equation~\cite{Lisa:2005dd}
\begin{equation}\label{eq:KooninPratt_Simple}
C(k)=\int S(\mt,r)\left|\Psi(\vec{k},\vec{r})\right|^2d^3r.
\end{equation}
Here, $C(k)$ represents the correlation as a function of the single-particle momentum $k$ in the pair rest frame, while $\Psi(\vec{k},\vec{r})$ is the wave function of the relative motion of the pair.
The source $S(\mt,r)$ is provided as a function of the relative distance between the particles $r$ at the moment of their effective emission.
The modeling of the source in small collision systems at the LHC has been extensively studied in several recent works~\cite{ALICE:Source,Mihaylov:2023pyl,ALICE:PionMax}, which provide compatible results on the source properties. In the present analysis we adopt the CECA model~\cite{Mihaylov:2023pyl}, which operates based on a common emission source for all primordial particles, accounting for particle production through the decay of short-lived resonances and incorporating an intrinsic \mt scaling of the source size, as observed in the data. The CECA framework utilizes three fit parameters, for details refer to~\cite{Mihaylov:2023pyl}.
For the ALICE data sample used in this work, the ``standard candle'' used to calibrate the source is the \pP correlation, which has been measured differentially in \mt~\cite{ALICE:Source,SupplementSource}. 
In the present analysis, we perform a pre-fit of the \pP correlations using the Argonne $v18$ potential~\cite{Wiringa:AV18}, following the same procedure as described in~\cite{Mihaylov:2023pyl}. However, out of the seven available \mt bins, we have omitted the last two due to issues with convergence at very low $k$.
To eliminate any bias related to the assumption of a common source, the extracted source parameters from the pre-fit of the \pP correlations are used as an initial guess for the \pL system, after which they are re-fitted alongside the six interaction parameters, allowing a variation of 3 standard deviations ($\sigma$). Finally, we verify that the parameters converge to a proper local minimum without reaching the limiting values. 
The remaining details on both the \pP and \pL fits, such as the inclusion of momentum resolution, feed-down, non-femtoscopic baseline, etc., are mirrored from the analysis of the same data described in~\cite{Mihaylov:2023pyl}.\\
\indent
The first objective of the present analysis is to quantify the allowed scattering lengths in the spin singlet/triplet channels ($f_{0/1}$), which can be accomplished by considering potentials of varying strengths in the CATS framework.
Note that we use the sign convention where attractive/repulsive interactions are characterized by positive/negative scattering lengths.
Both the cross section and the correlation function are composed of a weighted sum of the two channels, with respective weights of 1/4 and 3/4. The two spin states are attractive and exhibit similar correlation shapes that differ in magnitude. 
Due to this similarity, the present analysis is not particularly sensitive to the individual scattering lengths of each spin channel, and a unique solution is not expected. Nevertheless, requiring that the two-body forces alone produce a bound hypertriton puts a lower limit on the strength of the interaction 
in the spin singlet channel. A concrete estimate is difficult to provide,
however, judging from results for the hypertriton separation
energy from Faddeev calculations employing modern YN potentials
\cite{Haidenbauer:NLO19,Haidenbauer:NNLO}, 
values of $f_0\lesssim 2.0$~fm are not realistic. 
Assuming that the hypertriton is solely bound by three-body 
forces is likewise unrealistic given the present estimates
\cite{Le:2023bfj} and explicit calculations \cite{Kohno:2023xvh} 
of their possible contribution. 
In view of this the scan is performed for $f_0>1.6~$fm. 
The lack of a unique solution leads to convergence issues in the fit procedure. 
To address this problem, multiple fits are performed, each constrained within a specific small region of $f_0$ and $f_1$ values. The procedure is repeated until the entire desired phase space is scanned. The best $\chi^2$ of each individual step is saved, allowing the creation of an exclusion plot for $f_0$ and $f_1$. The estimator for the exclusion is the total $\chi^2 = \chi^2_\text{scattering}+\chi^2_\text{femtoscopy}$. The $\chi^2$ is converted into a number of standard deviations ($\mathrm{n}\sigma$) with respect to the best solution, accounting for a total of 9 degrees of freedom~\cite{NumericalRecipes}. 
\begin{figure*}[ht]
  \includegraphics[width=\textwidth]{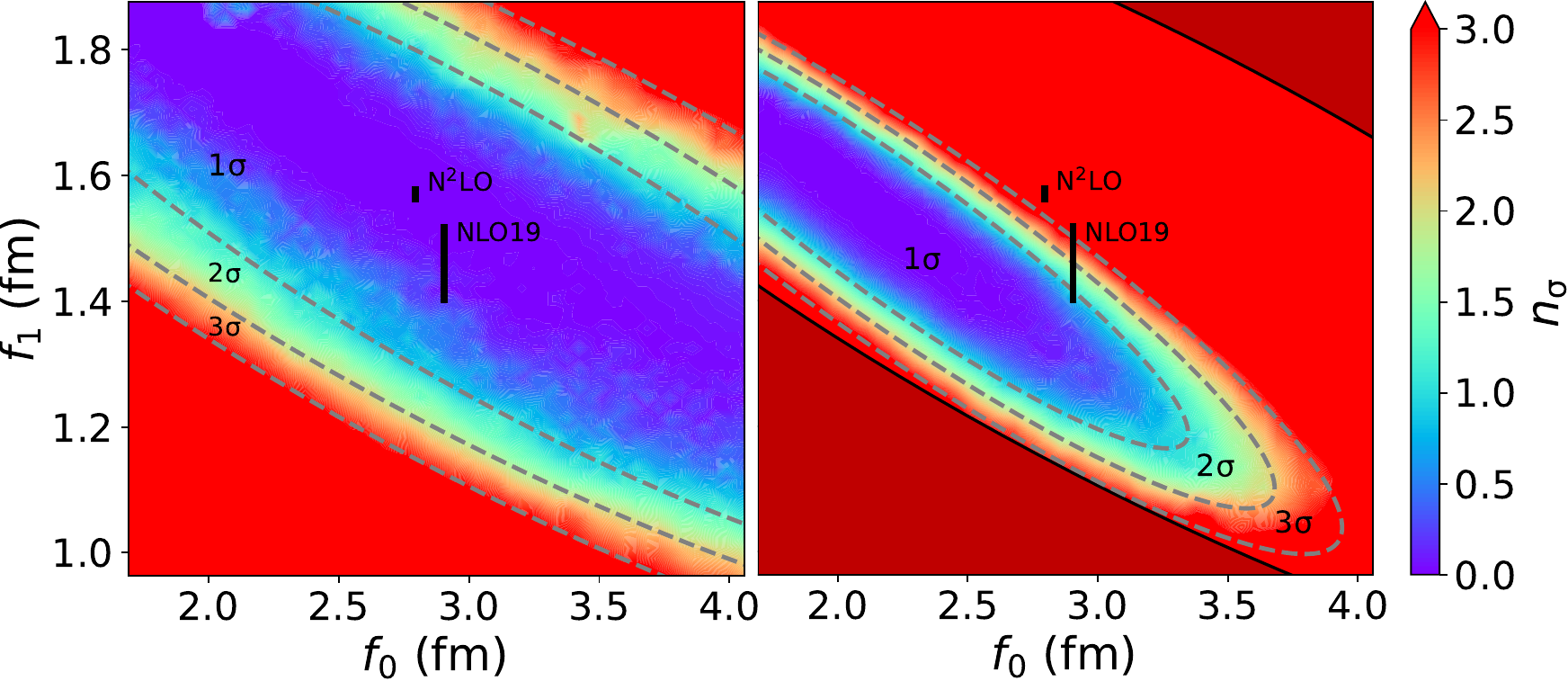}
  \caption{Exclusion plots for the singlet ($f_0$) and triplet ($f_1$) \pL scattering lengths based on the analysis of the cross section data (left panel) and on the combined analysis of cross section and correlation data (right panel). See text for details.}
  \label{fig:exclusion}
\end{figure*}

\newline\newline
\noindent
\textit{Results:}
The exclusion plot based on results with the Usmani potential is shown in Fig.~\ref{fig:exclusion}. The axes correspond to the scattering lengths in the singlet $f_0$ and triplet $f_1$ channel, while the color code contains information on the compatibility with the data. The left panel is based on the analysis of only the cross section, while the right panel is the final result based on the combined analysis of femtoscopic and scattering data. The gray dashed lines mark the 1, 2 and 
3$\sigma$ exclusion regions. The black solid line, in the right panel, marks the border of a 3$\sigma$ deviation with respect to the scattering data alone and is identical to the outer most dashed line from the left panel, while the shaded area depicts the region of even worse compatibility. As expected, there is a strong correlation between $f_0$ and $f_1$, and the inclusion of femtoscopy data into the analysis leads to a significant decrease in uncertainties. Values of $f_0>3.34~$fm or $f_1<1.18$~fm are disfavored by the data. The lower (upper) bound of $f_0$ ($f_1$) cannot be constrained within the investigated phase space. 
Figure~\ref{fig:exclusion} contains two vertical bars depicting 
the values of the scattering parameters based on the NLO19
\cite{Haidenbauer:NLO19} and 
the next-to-next-to-leading order N$^2$LO~\cite{Haidenbauer:NNLO} potentials. 
The size of the markers represents the uncertainties related to the employed regulator (cutoff $\Lambda$) in the chiral NY potentials. Both of these values are located approximately in the middle of the phase space region allowed by the scattering data alone, which is not surprising, as the LECs of those potentials have, up to now, been fitted 
to that data. Nevertheless, the enhanced sensitivity of the combined analysis shows that the predicted scattering lengths are disfavored by as much as 4.8$\sigma$ in the case of N$^2$LO. The NLO19 interaction is overall better in line with the present analysis, nevertheless, a systematic deviation of 
ca.~1-3$~\sigma$ is observed, depending on the cutoff value. 
Indeed, the predictions by the 
potential with cutoff $\Lambda=600$~\MeV of $f_0=2.91$~fm and
$f_1=1.41$~fm are in relatively good agreement, resulting in a deviation from the best solution of 1.1$\sigma$. On the other
hand, a best fit of $f_1$, keeping $f_0=2.91$~fm fixed, yields $f_1=1.32\pm0.08$~fm. Clearly, due to the strong correlation between the two parameters, changing the value of $f_0$ will 
influence the outcome for $f_1$. For example, fixing $f_0=2.1$~fm implies the value $f_1=1.56\pm0.11~$fm. Table~\ref{tab:ScattLengths} provides multiple examples for scattering parameters and their compatibility to the data. 
These results indicate an overall less attractive interaction compared to the published chiral potentials. 

\begin{figure}[h]
    \centering
    \includegraphics[width=0.48\textwidth]{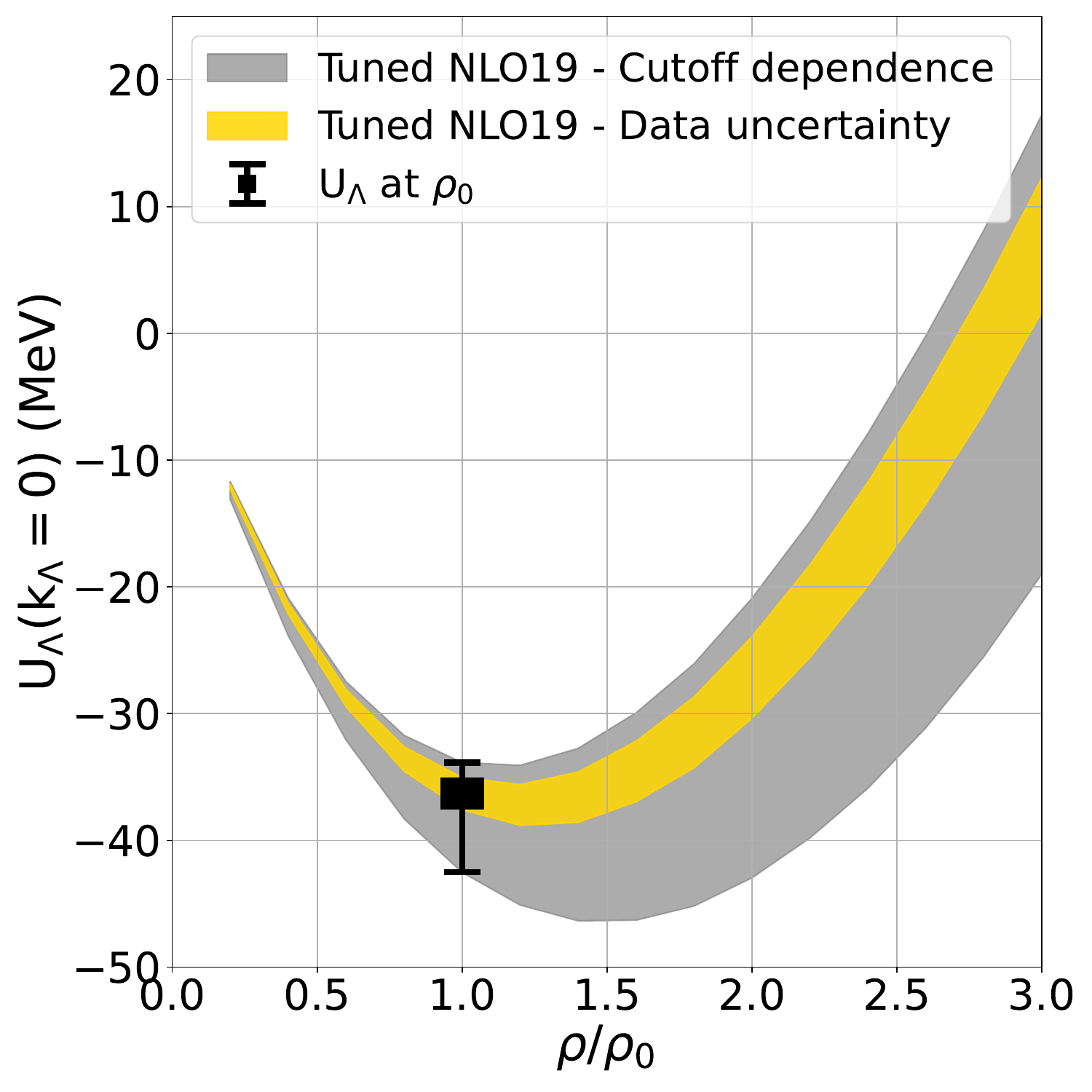}
    \caption[]{$\Lambda$ single-particle potential $U_\Lambda$ as a function of the nuclear-matter density $\rho$ with 
    $\rho_0 = 0.166$~fm$^{-3}$
    being the nuclear-matter saturation density. 
    }
    \label{fig:ulambda}
\end{figure}

\indent
As a next step we explore how this less attractive N$\Lambda$  
interaction affects predictions for the single-particle potential
$U_\Lambda$ at nuclear saturation density $\rho_0$ and 
its density dependence in general, considering 
the relevance of this quantity for the role of the $\Lambda$ hyperon 
in neutron stars~\cite{VidanaHyp,SchaffnerBielichBook}.
In Fig.~\ref{fig:ulambda},
we present results for the single-particle potential
$U_\Lambda(k_\Lambda=0)$ as a function of the nuclear matter density $\rho$, 
evaluated self-consistently within a conventional $G$-matrix calculation. We employ the formalism described in detail in
Refs.~\cite{Petschauer:2015nea,Gerstung:2020ktv}, 
where the so-called continuous choice is taken for the 
intermediate states, and the N$^3$LO potential from 
\cite{Entem:2003ft} is used for the NN interaction.
The NLO19(600) potential is chosen as starting
point from which, by readjusting 
some of its LECs to reproduce 8 combinations of 
the N$\Lambda $ spin
singlet and triplet scattering lengths within the 1$\sigma$ region (points i to viii in Tab.~\ref{tab:ScattLengths} in the Appendix), we estimate the corresponding $U_\Lambda(\rho_0)$ given by the black square in Fig.~\ref{fig:ulambda}.
The vertical error bar represents the theoretical uncertainty estimated by considering the NLO19 potentials with different cutoffs ($500-650$~MeV), all refitted to describe the same set of scattering parameters. 
The final result is  $U_\Lambda(\rho_0)= -36.3\pm 1.3 \mathrm{(stat)}^{+2.5}_{-6.2}\mathrm{(syst)}$~MeV, where the statistical uncertainty is associated with the data on the scattering parameters (right panel in Fig.~\ref{fig:exclusion}) and the systematic with the cutoff dependence from the theory.
The resulting theoretical uncertainty is large, as likewise reported in standard nuclear matter calculations with 
chiral nucleon-nucleon potentials~\cite{Sammarruca:2014zia,Hu:2016nkw}. 
Additionally, for the behavior of the potential depth as a function of $\rho$, we show in Fig.~\ref{fig:ulambda} the theoretical uncertainty (grey band) and the uncertainty from the combined data (yellow band). 
The predicted values for $U_\Lambda (\rho)$ are similar to the result for the
original NLO19 potential. Our 
results at $\rho_0$ lie below the usually 
cited semi-empirical value of $U_\Lambda = -28 \sim -30$~MeV obtained from hypernuclei constraints~\cite{GalReview}. This is in line with comparable
$G$-matrix calculations where a similar overbinding feature has
been observed when two-body-only contributions are taken into account,
as seen in, for example, \cite{Haidenbauer:NLO19,Rijken:1998yy}. 
In two recent works~\cite{Gerstung:2020ktv,Logoteta:2019utx} an overall repulsion from a chiral NNY three-body force has been incorporated in the form of an effective density-dependent
NY two-body potential \cite{Petschauer:2016pbn}. 
This addition allows the authors to satisfy the $\approx -30$ MeV hypernuclear constraint and, at the same time,
suppresses the appearance of $\Lambda$ hyperons for densities 
as realized in NS, i.e. offers a solution to the so-called
``hyperon-puzzle'' \cite{TolosFabbiettiReview,VidanaHyp}.
Our results are compatible with such a strategy. 
Quantitative constraints on this effective three-body force might become available in future correlation studies~\cite{Kievsky:2023maf,ALICE:2022boj}. 
\\ 
\indent
In conclusion, in this work we have presented the first combined analysis of low-energy femtoscopic and scattering data to constrain the $S$-wave scattering parameters of the \pL interaction, resulting in the tightest limits available for future theoretical studies. The p$\Lambda$ interaction is found to be
overall less attractive than what has been 
indicated by the scattering data from the 1960s. We observe a strong, approximately linear, correlation between the values of the scattering lengths in the spin singlet and triplet states. The best solution, if $f_0$ is fixed to 2.1~fm, is $f_1=1.56\pm0.08$~fm. Lower $f_0$ values will eventually prohibit a hypertriton bound by two-body forces. The maximum (minimum) allowed values for $f_0$ ($f_1$) are 3.34~(1.18)~fm.\\
\indent
Clearly, the accurate reproduction of low-energy
N$\Lambda$ data is an important requirement for realistic predictions of many-body systems.
Thus, we have fine-tuned the chiral YN potential NLO19 to match the established scattering parameters
in order to explore the impact on the in-medium 
properties of the $\Lambda$ hyperon.
The result with those NLO19 variants for $U_\Lambda(\rho_0)$ is $\sim -36$~MeV, which implies an overbinding with regard to the nominal $\Lambda$ binding energy in infinite nuclear matter. 
This is consistent with the current notion of an additional repulsion acting on the $\Lambda$ within the medium, attributed to many-body effects.
The presented results can serve as a state-of-the-art guideline for the contribution to be expected from two-body \pL interactions.

\begin{acknowledgements}
We would like to thank Prof. L. Fabbietti for her support and fruitful discussions which helped us in finalizing these results. 
This work was supported by the ORIGINS cluster DFG under Germany’s Excellence
Strategy - EXC2094 - 390783311 and the DFG through Grant SFB 1258 “Neutrinos and Dark Matter in Astro and Particle Physics”.
V. M. S. is supported by the Deutsche Forschungsgemeinschaft (DFG) through the grant MA $8660/1-1$.
\end{acknowledgements}

\section{Appendix}\label{sec:Supplement}

Table~\ref{tab:ScattLengths} presents several examples illustrating the compatibility of different scattering parameters with the data. The compatibility is estimated using the number of standard deviations, denoted as n$\sigma$, with respect to the best possible fit. We consider three scenarios: firstly, only the femtoscopy data is taken into account, resulting in n$\sigma_\text{fmt}$; secondly, only the scattering data is considered (n$\sigma_\text{sct}$); and thirdly, n$\sigma_\text{tot}$ corresponds to the combined analysis presented in this work. It is essential to note that these three scenarios represent independent analyses, each having a different best solution as a baseline. Consequently, there is no straightforward relation between the three estimators.

        \begin{table}[h]
        \caption{Summary table showing the compatibility of different scattering parameters to the femtocsopy data (n$\sigma_\text{fmt}$), scattering data (n$\sigma_\text{sct}$) as well as the combined analysis from the present work (n$\sigma_\text{tot}$).}
        \label{tab:ScattLengths}
        \begin{ruledtabular}
            \begin{tabular}{c|c|c|c|c|c}
            Usmani & $f_0$ (fm) & $f_1$ (fm) & n$\sigma_\text{fmt}$ & n$\sigma_\text{sct}$ & n$\sigma_\text{tot}$ \\
            parameterization & & & & & \\
            \hline
              NLO13(600) & 2.91 & 1.54 & 5.2 & 0.0 & 4.6 \\ \hline
      NLO19(600) & 2.91 & 1.41 & 1.7 & 0.4 & 1.1 \\ \hline
      N$^2$LO(550) & 2.79 & 1.58 & 5.4 & 0.0 & 4.8 \\ \hline
      i & 2.10 & 1.44 & 0.2 & 2.1 & 1.0 \\ \hline
      ii & 2.10 & 1.56 & 0.0 & 0.9 & 0.0 \\ \hline
      iii & 2.10 & 1.66 & 1.8 & 0.2 & 1.0 \\ \hline
      iv & 2.50 & 1.32 & 1.7 & 0.2 & 1.0 \\ \hline
      v & 2.50 & 1.46 & 0.2 & 0.8 & 0.0 \\ \hline
      vi & 2.50 & 1.55 & 1.8 & 0.2 & 1.0 \\ \hline
      vii & 2.91 & 1.32 & 0.1 & 1.5 & 0.3 \\ \hline
      viii & 3.34 & 1.18 & 1.2 & 0.9 & 1.0 \\ \hline

            \end{tabular}
        \end{ruledtabular}
        \end{table}

\label{Bibliography}
\bibliographystyle{unsrturl}
\bibliography{bibliography.bib}

\end{document}